\documentclass[10pt,
english,
parskip=half]{scrartcl}

\usepackage[latin1]{inputenc}
\usepackage{amsmath}

\usepackage{amsfonts}
\usepackage{physics}
\usepackage{amssymb}
\usepackage{faktor}
\usepackage{graphicx}
\usepackage[export]{adjustbox}
\usepackage{tikz}
\usepackage{amsthm}
\numberwithin{equation}{section}
\numberwithin{figure}{section}
\usepackage[english]{babel}
\usepackage{mathtools}
\usepackage{nicefrac}

\usepackage{tcolorbox}

\usepackage{csquotes}

\usepackage{dsfont}

\usepackage{a4} %A4 Support
\usepackage{a4wide}

\usepackage{subcaption}
\usepackage{abstract}
\usepackage{array}

\usepackage[backend=biber,autocite=footnote,style=ieee,citestyle=numeric-comp,sorting=none]{biblatex}
\newbibmacro{string+doiurlisbn}[1]{%
  \iffieldundef{doi}{%
    \iffieldundef{url}{%
      \iffieldundef{isbn}{%
        \iffieldundef{issn}{%
          #1%
        }{%
          \href{http://books.google.com/books?vid=ISSN\thefield{issn}}{#1}%
        }%
      }{%
        \href{http://books.google.com/books?vid=ISBN\thefield{isbn}}{#1}%
      }%
    }{%
      \href{\thefield{url}}{#1}%
    }%
  }{%
    \href{http://dx.doi.org/\thefield{doi}}{#1}%
  }%
}

\DeclareFieldFormat{title}{\usebibmacro{string+doiurlisbn}{\mkbibemph{#1}}}
\DeclareFieldFormat[article,incollection]{title}%
    {\usebibmacro{string+doiurlisbn}{\mkbibquote{#1}}}

\usepackage[]{authblk}

\usepackage[normalem]{ulem}
\newcommand{\FH}[1]{\textcolor{black}{#1}}
\newcommand{\TW}[1]{\textcolor{black}{#1}}

\newcommand{\WT}[1]{\textcolor{black}{#1}}

\usepackage[colorlinks=true,
linkcolor=black,
citecolor=black,
filecolor=black,
pagecolor=black,
urlcolor=magenta,
bookmarks=true,
bookmarksopen=true,
bookmarksopenlevel=3,
plainpages=false,
pdfpagelabels=true]{hyperref} 
\usepackage{cleveref}
\usepackage{upgreek}
\usepackage{array}%Erzeugen von Vektoren
\usepackage{bm}

\def \ba  {\begin{align}}
\def \ea  {\end{align}}

\author[1]{Torsten Weber}
\author[1]{Fabian Haneder}
\author[1]{Klaus Richter}
\author[1]{Juan Diego Urbina}
\affil[1]{Institut f\"ur Theoretische Physik, 
Universit\"at Regensburg, D-93040 Regensburg, Germany}

\title{Constraining  Weil-Petersson volumes by universal random matrix 
correlations in low-dimensional quantum gravity}
\makeindex
\begin{document}
\maketitle

\begin{abstract}
Based on the discovery of the duality between Jackiw-Teitelboim quantum gravity and a double-scaled matrix ensemble by Saad, Shenker and Stanford in 2019, we show how consistency between the two theories in the universal Random Matrix Theory (RMT) limit imposes a set of constraints on the volumes of moduli spaces of Riemannian manifolds. These volumes are given in terms of polynomial functions, the Weil-Petersson volumes, solving a celebrated nonlinear recursion formula that is notoriously difficult to analyze. Since our results imply  \textit {linear} relations between the coefficients of the Weil-Petersson volumes, they therefore provide both a stringent test for their symbolic calculation and a possible way of simplifying their construction. In this way, we propose a long-term program to improve the understanding of mathematically hard aspects concerning moduli spaces of hyperbolic manifolds by using universal RMT results as input.
\end{abstract}

\section{Introduction}
\setcounter{footnote}{0}
While initially, the methods and concepts of quantum chaos attempted to explain how chaos in a classical system finds its way into the observed universality of short-range spectral fluctuations in the corresponding quantised version \cite{Gutzwillerb,Haake2010a,Stoeckmannb}, since its precise formulation in the 1980's \cite{BGS} the connection between dynamical chaos and Random Matrix Theory (RMT) has also offered some deep insights into mathematical questions. A paradigmatic example of how such an approach works is number theory. Here, one starts from the conjectured RMT-like statistical distribution of the non-trivial zeroes of the Riemann zeta function \cite{Montgomery1973}, supported by a huge amount of numerical evidence \cite{Odlyzko1989}, and uses it to obtain number-theoretical results \cite{KeatingSnaith}, even defining a whole research program. 

The rationale of this approach can be exported to any field and regime of parameters where fidelity to RMT is expected to hold. Then, the universal -- and usually tractable -- RMT results can be considered as constraints, imposing relations between physical objects of the theory, similar to how the statistical correlations of prime numbers are constrained by RMT. This is particularly the case for theories where the microscopic mechanism responsible for classical chaotic dynamics is not well understood, and therefore the well-developed machinery of periodic orbit theory \cite{Gutzwillerb,Haake2010a,Berry1985a,Sieber2001,Muller2005,Richter2022} cannot be invoked to explain RMT-like features.

Remarkably, certain aspects of quantum gravity fall into this category. This is because, although the precise connection between periodic orbit theory and the conjectured chaotic character of important quantum gravitational models \cite{Sekino2008,Shenker2014a,Maldacena2016,Cotler2017,Saad2018,Susskind2021a} is still an open problem, an exact mapping between 2D dilaton gravity (so-called Jackiw-Teitelboim gravity \cite{Jackiw1985,Teitelboim1983}) and a matrix model \cite{Saad2019} has recently been discovered, creating an explosion of interest \cite{Stanford2019,Saad2019a,Garcia-Garcia2020,Okuyama2020,Janssen2021}. 

The present paper (and its companion \cite{BlommaertAndFriends}) follows the route depicted above. We invoke the well justified \textit{assumption} that in a certain precise limit, spectral correlations computed from the exact solution of JT gravity are identically given by the universal RMT results. Imposing such an equivalence then constrains the objects appearing on the JT side, which in this case turn out to be related to the moduli space of two-dimensional manifolds. Our objective is to make these constraints explicit, and to initiate the study of their structure and consequences. 
Doing so, we provide further evidence for the equivalence of JT gravity and universal RMT, and in particular the operational meaning of the ``universal limit'' as proposed in \cite{Saad2019a} and further developed in \cite{StanfordAndFriends}.

\begin{figure}[h]
    \centering
    \begin{displaymath}
    \ev{Z(\beta)}=\includegraphics[width=0.9\textwidth,valign=c]{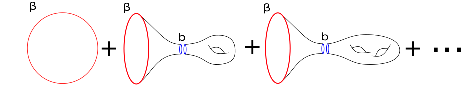}
    \end{displaymath}
    \caption{\TW{Example of the genus expansion in \cref{eq:genus_expansion} \FH{for the case $n=1$}. The asymptotic AdS boundary of length $\beta$ is depicted in red, while the geodesic boundary of length b along which the manifold is cut is depicted in blue.}}
    \label{fig:1point_cut}
\end{figure}

To begin with, we now briefly present the main features of Jackiw-Teitelboim gravity, which is a two-dimensional theory of gravity coupled to a dilaton $\phi$, described by the action 
\begin{equation}
    S_{\text{JT}}=-\frac{S_0}{2\pi}\chi(\mathcal{M})-\frac{1}{2}\int_\mathcal{M}\sqrt{\FH{\det g_{\mu\nu}}}\phi(R+2)-\int_{\partial\mathcal{M}}\sqrt{h}\phi(K-1), \label{eq:action}
\end{equation}
\FH{on a manifold $\mathcal{M}$ with boundary $\partial\mathcal{M}$, where $g_{\mu\nu}$ is a metric on $\mathcal{M}$, $R$ the Ricci scalar computed from $g_{\mu\nu}$, $h$ the induced metric on $\partial\mathcal{M}$, and $K$ the extrinsic curvature.} 
The first term is the Euler characteristic of the manifold $\mathcal{M}$, multiplied by a \FH{large} constant $S_0$, \FH{chosen such that $e^{S_0}$ is a characteristic scale of the density of states. This term} causes the path integral over \eqref{eq:action} to decompose into a genus expansion of the form\footnote{\FH{The notation $\ev{Z(\beta_1)\ldots Z(\beta_n)}$ for these partition functions is chosen to emphasize the duality with correlation functions of the heat kernel in the matrix model introduced later, but it can be made more rigorous by defining a boundary creation operator in the spirit of \cite{Marolf2020} and inserting it in the path integral.}}
\begin{equation}
    \ev{Z(\beta_1)\ldots Z(\beta_n)}=\sum_{g=0}^{\infty}e^{(2-2g-n)S_0}\int_0^\infty b_1\dd{b_1}\ldots b_n\dd{b_n}
    Z_\text{tr}(\beta_1,b_1)\ldots Z_\text{tr}(\beta_n,b_n)V_{g,n}(b_1,\ldots,b_n)\label{eq:genus_expansion}
\end{equation}
\FH{for $n$ asymptotically AdS boundaries of length $\beta_j$. Here,} $Z_\text{tr}(\beta,b)\WT{=\sqrt{\frac{\gamma}{2\pi \beta}}e^{-\frac{\gamma b^2}{2\beta}}}$ is \FH{the partition function of a hyperbolic ``trumpet'' \TW{(cf. \cref{fig:1point_cut})} with one asymptotically AdS boundary of length $\beta$ and one geodesic boundary of length $b$} (the precise form of which is not important here), and $V_{g,n}(b_1,\ldots,b_n)$ are the Weil-Petersson volumes, that will be considered in more detail later on. \FH{They arise from the bulk part of the path integral and describe the volume of the moduli space of each topological sector labelled by $g,n$}\TW{, that depends on the boundary lengths $\qty(b_1,\ldots b_n)$. A good way to visualize the individual terms appearing in \cref{eq:genus_expansion}, for the simplest example of only one asymptotic AdS boundary of length $\beta$, is shown in \cref{fig:1point_cut}, where the contributions to the partition function to the genera 0, 1 and 2 are depicted. The splitting of both non-zero genus contributions into the boundary part (``trumpet'') and the bulk part, carrying the genus illustrates the way they are computed in \cref{eq:genus_expansion} and actually corresponds to the way one proves this formula via the computation of the path integral \cite{Saad2019,Stanford2019}.} Eq. \eqref{eq:genus_expansion} holds except for special cases\footnote{The $n=1$ result is the disk $Z_\text{disk}(\beta)=e^{S_0}\frac{\gamma^{\frac{3}{2}}e^{\frac{2\pi^2\gamma}{\beta}}}{(2\pi)^{\frac{1}{2}}\beta^{\frac{3}{2}}}$, while for $n=2$, one has the ``double-trumpet'', which can be written as part of \eqref{eq:genus_expansion} by formally defining $V_{0,2}(b_1,b_2)=\delta(b_1^2-b_2^2)$.}, the genus 0 contributions for $n=1,2$. 

This theory has received a lot of attention in recent years after a duality between JT gravity and a certain double-scaled Hermitian matrix model was established in \cite{Saad2019}. To be more precise, the authors of Ref.~\cite{Saad2019} found a \WT{(formal)} matrix model\WT{\footnote{\WT{In the following we will refer to this as a ``matrix model'' for brevity and refer to \cref{app:top_recursion} to an explanation of why it is more precisely denoted as ``formal''.}}} defined by some potential $V(H)$, and performed the limit $\dim H=N\to\infty$ in such a way that the leading spectral density remained normalised \footnotemark, yielding
\begin{equation}\label{eq:rho_0_t}
    \rho_0^T(E)=e^{S_0}\frac{\gamma}{2\pi^2}\sinh(2\pi\sqrt{2\gamma E})\FH{,}
\end{equation}
\WT{\footnotetext{\WT{In the sense of double-scaled matrix models. Briefly put this means that prior to taking the limit $\dim H=N\to\infty$ the leading order spectral density has compact support on $[0,a]$ for $a\in\mathbb R$ and is bounded on this interval, thus $C(a)\coloneqq \int_{\mathbb{R}} \rho_0$ is finite. This is done such that the $\sinh$ behaviour is obtained in the limit of large $a$. Taking the} \WT{double-scaling limit means taking the limit $a\to \infty$, thus $C(a)\to \infty$, simultaneously with $N\to\infty$ while keeping $e^{S_0}\coloneqq \frac{N}{C(a)}$ constant. Thus, the spectral density is, strictly speaking, not normalisable but is finite and controlled by $e^{S_0}$ for not infinite positive real numbers which, as it will turn out later, is the only region relevant for the computation of correlation functions. A more detailed explanation can be found in \cite{Saad2019,Eynard2007b}. For an explanation using the connection with the SYK model see e.g. \cite{Cotler2017}.}}}\FH{where $\gamma$ is the characteristic energy scale of the system, usually set to $1/2$. The subscript 0 refers to the genus 0 part of the eigenvalue density, while the superscript $T$ (for ``total'') indicates that the density (before taking the limit $N\to\infty$) is normalised to $N$, rather than to 1. $\rho^T=e^{S_0}\rho$ for the eigenvalue density $\rho$ used below.}
In the course of taking this limit, the usual perturbative $1/N$ expansion of the matrix model is replaced by an expansion in $e^{-S_0}$, e.g. for correlators of the partition function\footnote{\FH{One can compute the correlation functions of $\rho^T(E)$ as inverse Laplace transforms of the correlators of $Z(\beta)$. Hence, the spectral density correlators exhibit a genus expansion of the same form as \eqref{eq:genus_expansion_MM}.}} $Z(\beta)=\tr e^{-\beta H}$, 
\begin{equation}\label{eq:genus_expansion_MM}
    \ev{Z(\beta_1)\ldots Z(\beta_n)}=\sum_{g=0}^{\infty}e^{(2-2g-n)S_0}Z_{g,n}(\beta_1,\ldots,\beta_2).
\end{equation}
It is this genus expansion that has been shown to exactly compute \eqref{eq:genus_expansion}. However, establishing the translation of nonperturbative aspects of the matrix model to JT gravity has proven more challenging (see however \cite{Blommaert2019,Okuyama2021b,Griguolo2021,Gao2021a}). A key reason for this is that the matrix model of \cite{Saad2019} suffers from a nonperturbative instability, meaning that the integration contour of the matrix integral must be deformed. This process is not unique, however, and thus leads to an ambiguity in the nonperturbative completion. Much work has been done on trying to give JT gravity a rigorous nonperturbative definition and to study its features, particularly by Johnson \cite{Johnson2020,Johnson2020a,Johnson2020b,Johnson2020c,Johnson2021,Johnson2021a,Johnson2021c,Johnson2021d,Johnson2022,Johnson2022wsr}, and more recently, a nonperturbative completion in terms of Kodaira-Spencer theory has been given \cite{Post2022,Altland2022}.

In light of the amount of interest particularly the nonperturbative sector of the JT matrix model has received, it seems useful to look at concrete nonperturbative features of the matrix model and study how they are realised in JT gravity. Particularly, we want to focus on what we call the universal limit of the matrix model\footnote{We sometimes refer to this limit as the RMT limit, or particularly when talking about JT, the $\tau$-scaling limit.} (see e.g. \cite{Haake2010a}). This is the limit of the matrix model in which correlation functions are given by the ones obtained in the appropriate Altland-Zirnbauer ensemble (in the present case, the Gaussian unitary ensemble). We will describe how to access this limit in JT gravity below, but for now it is sufficient to recall that in this limit, correlation functions of e.g. the level density are described by universal, \textit{finite} functions. From this latter property alone, we will be able to derive very nontrivial identities between coefficients of the \TW{Weil-Petersson volumes}, i.e. the polynomials $V_{g,n}$ appearing in \eqref{eq:genus_expansion}. 

The Weil-Petersson (WP) volumes \cite{Mirzakhani2007} describe the volume of the moduli space of hyperbolic surfaces with genus $g$ and $n$ geodesic boundaries with lengths $b_1,\ldots,b_n$. In principle, they are computable individually from scratch by performing all possible decompositions of the manifold in terms of three-holed spheres, parametrising these decompositions by Fenchel-Nielsen coordinates $(l_i,\tau_i)$ and integrating the Weil-Petersson form in these coordinates $\omega_\text{WP}=dl_i\wedge d\tau_i$ over the moduli space, while modding out the mapping class group of the surface to account for overcounting the decompositions. 

In practice, however, this is not feasible and one uses Mirzakhani's recursion relation \cite{Mirzakhani2007} satisfied by the $V_{g,n}$ instead. While this is doable, the effort required to determine the volumes particularly for high $g$ or $n$ is still substantial, and a method of relating the different terms in the WP volumes might be useful to simplify their calculation. 

In this paper then, we will show that the existence of a finite universal limit of the JT matrix model implies constraints on the coefficients appearing in the WP volumes $V_{g,2}(b_1,b_2)$ determining the JT gravity 2-point function. To this end, we will begin in \cref{sec:ansatz} by computing the spectral form factor in the universal limit, using the full, non-perturbative result for the spectral two-point function. Then, we will redo the calculation in \cref{sec:scaling} using the genus expansion, and compare the two results. Demanding that they are at least compatible (i.e. that the perturbative result does not diverge more strongly than the nonperturbative one) will yield constraints relating certain coefficients appearing in a given $V_{g,2}$. Finally, in \cref{sec:discussion}, we discuss the possible relation of our results to other areas of research. 
\TW{In \cref{app:proof} we provide a proof for the simplification of the constraints mentioned above. In \cref{app:g_5} we check the constraints for the case of $g=5$. In \cref{app:top_recursion} we give a review of the application of the topological recursion to the setting of JT gravity with the aim of showing how we obtained the WP volumes used for checking the cancellations we predict. A collection of these can be found in \cref{app:WP}.}
As a last note, the same cancellations as the ones we report have been found independently using intersection theory computations in \cite{BlommaertAndFriends}.

\section{The late time spectral form factor from random matrix universality}\label{sec:ansatz}
In this section, we will compute the late time spectral form factor (SFF) of the matrix model of \cite{Saad2019}\FH{, $\ev{Z(\beta+it)Z(\beta-it)}$.} \TW{In order to do \FH{so,} we will first recall \FH{some necessary} information on this matrix model. As already \FH{mentioned} in the introduction, the matrix model of \cite{Saad2019} is \FH{defined by specifying the} symmetry class (\FH{unitary,} i.e. Hermitian matrices) and giving the leading order density of states as 
\begin{align}\label{eq:rho_E}
    \rho_0(E)=\frac{\gamma}{2\pi^2}\sinh(2\pi\sqrt{2\gamma E}).
\end{align}
Note, that this density exhibits a different behaviour than a conventional one-cut model, e.g. the semicircle law obtain\FH{ed} for the Gaussian model as it does not have two spectral edges but only \FH{the} one at $E=0$. This is due to the double-scaling procedure needed to obtain this model from a ``standard'' (i.e. compact support of $\rho_0$) matrix model, outlined in the introduction. Furthermore, it suffices to give the leading order density of states as all orders of the perturbative expansion\FH{s of e.g. partition function correlators} in the small parameter of the double-scaled theory $e^{S_0}$ are \FH{determined by} the Eynard-Orantin topological recursion \cite{Eynard2004}. For the sake of completeness, we include the relevant definitions and formulae in \cref{app:top_recursion}. \FH{As already mentioned in the introduction, the aforementioned perturbative expansions can be evaluated and found to be in agreement with the JT gravity result \eqref{eq:genus_expansion}.} This is the main result of \cite{Saad2019} and, as a (pleasant) side effect, enables a \FH{simplified} computation of the WP volumes, \FH{since} the topological recursion is easier to solve than Mirzakhani's recursion. As the spectral form factor is defined as a correlator of partition functions, one can \FH{now} compute it to all orders in perturbation theory using the topological recursion. This, however, is not special \FH{and can be done} for all correlators. What is special about the SFF is that it}
\FH{has a universal form for chaotic systems. \TW{It is} given by a linear ramp region up to a large scale called the Heisenberg time, followed by a plateau. This universal form is identical to the result computed from Random Matrix Theory, though it cannot be directly observed in a single chaotic system, as it would display large-scale erratic oscillations. Rather, one has to perform some kind of average (e.g. the disorder average typically used in the SYK model \cite{Saad2018}). Of particular interest for us is the  plateau region, since it is a nonperturbative effect in RMT, and thus should naively not be captured by the JT gravity genus expansion. Finding a plateau is crucial for a theory of a quantum black hole however, since it allows the conclusion that the spectrum of the black hole is discrete \cite{Berry1985a}. Interestingly, as we will see below, a careful computation of the late time SFF of the matrix model of \cite{Saad2019} reveals a function of the time that exhibits the plateau, but also allows us to predict relations between coefficients in the Weil-Petersson volumes}. To do so, recall that the two-point function of \(Z(\beta)=\tr e^{-\beta H}\) can be expressed as a Laplace transform,
\begin{equation}
\TW{\ev{ Z(\beta_1)Z(\beta_2)}}=\int_{-\infty}^{\infty}\int_{-\infty}^{\infty}\ev{\rho^T(E_1)\rho^T(E_2)} e^{-\beta_1 E_1-\beta_2 E_2}\dd{E_1}\dd{E_2},
\end{equation}
where \(\rho^T(E)\) is the \textit{total} spectral density in the convention of \cite{Saad2019}, i.e. it is of order $e^{S_0}$\TW{(cf. \cref{eq:rho_0_t})}\footnote{\FH{This is taken to mean that the leading term in the perturbative expansion of the one-point function of $\rho^T$ is of order $e^{S_0}$.}}. This yields a useful form for the SFF,
\begin{equation}
\FH{\ev{Z(\beta+it)Z(\beta-it)}}\eqqcolon\kappa_{\beta}(t)=\int_{-\infty}^{\infty}\dd E e^{-2\beta E}\int_{-\infty}^{\infty}\dd{\Delta}e^{-it\Delta}\ev{\rho^T\qty(E+\frac\Delta 2)\rho^T\qty(E-\frac\Delta 2)},
\label{eq:kappa(tau)}
\end{equation}
\FH{where we substituted $E_{1/2}=E\pm\Delta/2$.} Next, we study the universal behaviour of \(\kappa_{\beta}(t)\) by evaluating it at times scaling with the Heisenberg time $e^{S_0}$. We define a rescaled time
\begin{equation}
	\tau:=e^{-S_0}t,
\end{equation}
and obtain
\begin{equation}
	\begin{aligned}
	\kappa_{\beta}(\tau)&=\int_{-\infty}^{\infty}\dd E e^{-2\beta E}\int_{-\infty}^{\infty}\dd{\Delta}e^{-ie^{S_0}\tau\Delta}\ev{\rho^T\qty(E+\frac\Delta 2)\rho^T\qty(E-\frac\Delta 2)}\\
	&=e^{2S_0}\int_{-\infty}^{\infty}\dd E e^{-2\beta E}\int_{-\infty}^{\infty}\dd{\Delta}e^{-ie^{S_0}\tau\Delta}\ev{\rho\qty(E+\frac{\Delta}{2})\rho\qty(E-\frac{\Delta}{2})},
	\end{aligned}
\end{equation}
making the dependence on $e^{S_0}$ explicit in the second line. The rapidly oscillating factor $e^{-ie^{S_0}\tau\Delta}$ localises the integral near small $\Delta$ (more precisely, of order \(e^{-S_0}\)), making it sufficient to evaluate $\ev{\rho(E+\frac{\Delta}{2})\rho(E-\frac{\Delta}{2})}$ for small differences in the arguments. 

To find an expression for this, we utilise the universal RMT limit of the matrix model. \TW{Note that it is not possible to take the ``usual'' universal limit $N\to\infty$, since this has already been performed in the course of implementing the double scaling limit, where $N$ is replaced by $e^{S_0}$. Hence, we identify the appropriate universal limit for the double scaled theory as $e^{S_0}\to\infty$, and adapt the expressions in \cite{Haake2010a} accordingly.} Note that we leave the firm ground established in \cite{Saad2019} here, since the universal limit of RMT correlation functions is a nonperturbative result, while the duality uncovered in \cite{Saad2019} is only at the perturbative level\footnote{A nonperturbative completion of this kind can however be interpreted in the context of minimal string theory in terms of D-brane insertions \cite{Saad2019,Blommaert2021}. In particular, this is consistent with the nonperturbative completion in terms of Kodaira-Spencer universe field theory \cite{Altland2022,Post2022}. However, our computation does not rely on a particular (stringy) UV completion of JT gravity or a specific nonperturbative matrix model, but only on the existence of a perturbatively dual matrix model of GUE symmetry class.}. To finally express the universal limit, we introduce a different, useful notation \cite{Haake2010a},
\begin{align}
&R_1(E)=e^{S_0}\ev{\rho(E)}\\
&R_2(E_1,E_2)=e^{2S_0}\ev{\rho(E_1)\rho(E_2)}-e^{S_0}\delta(E_1-E_2)\ev{\rho(E_1)}.\label{eq:universal}
\end{align}
Accordingly, the connected two-point correlation function is given by
\begin{align}
C_2(E_1,E_2)&=R_1(E_1)R_{\FH{1}}(E_2)-R_2(E_1,E_2)\\
&=e^{2S_0}\ev{\rho(E_1)}\ev{\rho(E_2)}+e^{S_0}\delta(E_1-E_2)\ev{\rho(E_1)}-e^{2S_0}\ev{\rho(E_1)\rho(E_2)}.
\end{align}
The universal limit, which is valid for all matrix models of unitary symmetry class, now takes the form \FH{\cite{Haake2010a}}\footnote{\TW{\FH{Note} that this result does not hold at the edge of the spectrum, i.e. $E=0$. \FH{However, consider the main result of the current section, \cref{eq:StanfordAnsatzShort} for the Airy model defined by $\rho_0(E)=\frac{1}{2\pi} \sqrt{E}$, as elaborated in \cite{StanfordAndFriends}, which reflects only the behaviour near the spectral edge of the JT gravity matrix model. An exact result for the SFF is available in this case, and by comparison, one finds that the ``sine kernel result'' \eqref{eq:StanfordAnsatzShort} reproduces only the leading term of the exact result, failing to capture corrections that are (crucially) subleading in $e^{S_0}$. For the present argument, we only require the SFF to scale \emph{no more strongly than} $e^{S_0}$, whence it is} sufficient to consider the contributions coming from the sine kernel. \FH{W}e re\FH{f}rain from \FH{performing} a more in-depth \FH{discussion} in this work, \FH{and refer the interested reader to the recent examination} of the corrections to the sine kernel arising from the more general Christoffel-Darboux kernel in \cite{Okuyama2023}, \FH{who show} that corrections to the sine kernel in \FH{the} general setting of topological gravity are indeed subleading, \FH{providing further} evidence \FH{for} the point we wanted to make plausible here.}}
\begin{align}
	\lim\limits_{e^{S_0}\to\infty}\frac{C_2(E_1,E_2)}{R_1(E_1)R_{\FH{1}}(E_2)}=\frac{\sin[2](\pi\ev{\rho(E)} e^{S_0}\Delta)}{\pi^2\ev{\rho(E)}^2e^{2S_0}\Delta^2},
\end{align}
where $e^{S_0}\Delta:=\tilde{\Delta}$ is kept finite in accordance with the above requirement of evaluating the correlation function for energy difference of the order of the inverse Heisenberg time\footnote{Note that the same limit is considered, and the same universal result used in \cite{Iliesiu2021}. However, the cancellation they observe for the 2-point function between perturbative and nonperturbative divergent pieces is not the same as the cancellation we will observe. In particular, the disconnected term in \eqref{eq:universal} vanishes on its own in the SFF when taking the limit.}. This also justifies the replacement $\ev{\rho(E_1)}\approx\ev{\rho(E_2)}\approx\ev{\rho(E)}$. Using the explicit expressions and taking the universal limit, we can solve for the desired correlation function,
\begin{align}
	\ev{\rho\qty(E+\frac{\Delta}{2})\rho\qty(E-\frac{\Delta}{2})}=\ev{\rho(E_1)}\ev{\rho(E_2)}+\delta\qty(\tilde{\Delta})\ev{\rho(E)}-\frac{\sin[2](\pi\ev{\rho(E)} \tilde{\Delta})}{\pi^2\tilde{\Delta}^2}.
\end{align}
We can now plug this back into the SFF \cref{eq:kappa(tau)} to find
\begin{equation}
    \begin{aligned}
	\kappa_{\beta}(\tau)=&\ev{ Z(\beta+ie^{S_0}\tau)}\ev{ Z\qty(\beta-ie^{S_0}\tau)}\\
	&+ e^{S_0}\int_{-\infty}^{\infty}\dd E e^{-2\beta E}\int_{-\infty}^{\infty}\dd{\tilde{\Delta}}
	e^{-i\tau\tilde{\Delta}}\qty[\delta\qty(\tilde{\Delta})\ev{\rho(E)}-\frac{\sin[2](\pi\ev{\rho(E)} \tilde{\Delta})}{\pi^2\tilde{\Delta}^2}]\\
	=&\ev{ Z(\beta+ie^{S_0}\tau)}\ev{ Z\qty(\beta-ie^{S_0}\tau)}+e^{S_0}\int_{E_0}^{\infty}\dd E e^{-2\beta E}\min\qty{\frac{\tau}{2\pi},\ev{\rho(E)}},
	\end{aligned}\label{eq:StanfordAnsatz}
\end{equation}
with $E_0$ the left edge of the spectrum of $\rho(E)$. 
At this point, a comment on our notion of taking the limit is in order. Although taken to infinity in the strict universal limit, $e^{S_0}$ is still around. This can be understood in the sense of isolating the most strongly divergent term in the limit $e^{S_0}\to\infty$. \TW{Considering in the same spirit the first term of \cref{eq:StanfordAnsatz} for the $g=0$ contribution to this product of correlators, one finds
\begin{equation}
	\begin{aligned}
	Z_0(\beta+ie^{S_0}\tau)Z_0(\beta-ie^{S_0}\tau)&=\frac{1}{\sqrt{2\pi}}\left(\frac{\gamma}{\beta+ie^{S_0}\tau}\right)^{\frac 32}e^{\frac{2\gamma\pi^2}{\beta+ie^{S_0}\tau}}\frac{1}{\sqrt{2\pi}}\left(\frac{\gamma}{\beta-i\tau}\right)^{\frac 32}e^{\frac{2\gamma\pi^2}{\beta-i\tau}}\\
	&=\frac{\gamma^3}{2\pi}\qty(\frac{1}{\beta^2+e^{2S_0}\tau^2})^{\frac{3}{2}}e^{2\gamma\pi^2\frac{2\beta}{\beta^2+e^{2S_0}\tau^2}},
	\end{aligned}
\end{equation}
which is subleading in $e^{S_0}$\FH{, even after} including the additional factor of $e^{S_0}$ one has to multiply to compute $\langle Z(\beta+it)\rangle$. Due to the suppression in powers of $e^{S_0}$ apparent from \cref{eq:genus_expansion}, all higher genus terms are subleading as well.
Likewise, it is enough to consider the genus 0 part of $\ev{\rho(E)}$, given by $e^{-S_0}\rho_0^T(E)$.} With these simplifications, it is possible to give an explicit result for the JT gravity SFF:

\begin{equation}
\kappa_{\beta}(\tau)=e^{S_0}\int_{0}^{\infty}\dd E e^{-2\beta E}\min\qty{\frac{\tau}{2\pi},\rho_0(E)}.
\label{eq:StanfordAnsatzShort}
\end{equation}
This result has been first reported in \cite{Saad2019a} and evaluated in \cite{StanfordAndFriends}. Important for us is that $\kappa_\beta(\tau)$ is finite at finite $\tau$.

\section{Scaling: Power counting considerations}\label{sec:scaling}
In the previous section, we derived an expression for the late time SFF of JT gravity from nonperturbative information about the dual matrix model. Another way to compute this quantity would be to use the JT genus expansion (which is equivalent to the perturbative expansion of the matrix model) and then perform the universal and late time (``$\tau$-scaling'') limit $e^{S_0}\to\infty,\,t\to\infty,\,\tau=e^{-S_0}t=\text{const.}$

In this limit, the SFF (as calculated from the genus expansion) takes the form 
\begin{equation}
    e^{S_0}\left(a_1\tau+a_3\tau^3+a_5(\beta)\tau^5+\ldots\right),\label{eq:powerSeries}
\end{equation}
with coefficients independent of $e^{S_0}$. The $a_i(\beta)$ specify a convergent series\footnote{Technically, a series with finite radius of convergence, which can be smoothly continued to $\tau\to\infty$.} which, importantly, agrees with the Taylor expansion of the result \eqref{eq:StanfordAnsatzShort} around $\tau=0$ and moreover gives the exact SFF for JT gravity in the $\tau$-scaling limit, including the plateau \cite{StanfordAndFriends}.

In order to systematically compute the contributions to \eqref{eq:powerSeries}, we employ the polynomial structure of the WP volumes \cite{Mirzakhani2007}:
\begin{equation}\label{eq:def_WP}
	V_{g,2}(b_1,b_2)=\sum_{\FH{n,m}}^{\FH{n+m}\leq 3g-1}C^{(g)}_{\FH{n,m}}b_1^{2\FH{n}}b_2^{2\FH{m}}
\end{equation}
with non-negative constants $C^{\WT{(g)}}_{\FH{n,m}}\in \pi^{6g-2-2\FH{(n+m)}}\cdot\mathbb{Q}$.
\TW{Using this, one can write the genus $g$ contribution to $\expval{Z(\beta_1)Z(\beta_2)}$ for arbitrary complex temperatures $\beta_1,\beta_2$
\begin{equation}
\begin{aligned}
    \expval{Z(\beta_1)Z(\beta_2)}&=\sum_{g=0}^\infty e^{-2g S_0}\int_{[0,\infty]^2}b_1\dd{b_1}b_2\dd{b_2}Z_{\text{tr}}(\beta_1,b_1)Z_{\text{tr}}(\beta_2,b_2)\sum_{n,m}^{n+m\leq 3g-1}C^{(g)}_{n,m}b_1^{2n}b_2^{2m}\\
    &=\sum_{g=0}^\infty e^{-2gS_0}\sum_{n,m}^{n+m\leq 3g-1}C^{\WT{(g)}}_{n,m}  \frac{\sqrt{{\beta}_1 {\beta}_2}}{\pi }\frac{n!m!2^{n+m-1}}{\gamma^{n+m+1}}{\beta}_1^n{\beta}_2^m.
\end{aligned}
\end{equation}
Thus, for the spectral} \WT{form} \TW{factor (i.e. putting $\beta_1=\beta+it$, $\beta_2=\beta_1^*$), one finds 
\begin{equation}\label{eq:SFF_t_expansion}
    \begin{aligned}
        \kappa_\beta(t)&=\sum_{g=0}^\infty e^{-2gS_0}\sum_{n,m}^{n+m\leq 3g-1}C^{\WT{(g)}}_{n,m} \frac{\sqrt{\beta^2+t^2}}{\pi }\frac{n!m!2^{n+m-1}}{\gamma^{n+m+1}}\qty(\beta+it)^n\qty(\qty(\beta+it)^*)^m\\
        &=\sum_{g=0}^\infty e^{-2gS_0}\sum_{n,m}^{n+m\leq 3g-1}C^{\WT{(g)}}_{n,m}  \frac{\sqrt{\beta^2+t^2}}{\pi }
		\frac{n!m!2^{n+m-1}}{\gamma^{n+m+1}}\sum_{k=0}^{n}\sum_{j=0}^{m}
		\binom{m}{j}\binom{n}{k}i^{j+k}\qty(-1)^{j}\beta^{m+n-k-j}t^{j+k}\\
    &\eqqcolon \sum_{g=0}^\infty e^{-2gS_0} \sum_{n,m}^{n+m\leq 3g-1}C^{\WT{(g)}}_{n,m}\qty(\kappa_{\beta}^g(t))^{n,m},
    \end{aligned}
\end{equation}
where in the second to last line the binomial theorem has been used to expand the powers of complex temperatures.
}
We now consider \TW{the contribution} \FH{to \eqref{eq:SFF_t_expansion}} for some fixed $g$ in the $\tau$-scaling limit. For given $(n,m)$, take for example the leading term \TW{in $t$ of $\qty(\kappa_{\beta}^g(t))^{n,m}$}, scaling like
\begin{equation}
    t^{n+m+1}=e^{(n+m+1)S_0}\tau^{n+m+1}.
\end{equation}
Meanwhile, \TW{the contribution to the spectral form factor of each} $\qty(\kappa_{\beta}^g(t))^{n,m}$ is suppressed by a factor $e^{-2gS_0}$ due to the genus expansion. As an example, the highest order term coming from $n+m=3g-1$ contributes to the SFF to the order
\begin{align}
	e^{-2gS_0}e^{3gS_0}\tau^{3g}=e^{gS_0}\tau^{3g}.
\end{align}
The universal part of the SFF emerges by performing the $\tau$-scaling limit and extracting the leading term, which should be of order $e^{S_0}$. Clearly however, there are contributions to the $g>1$ SFF (or rather, to $e^{-S_0}\mathrm{SFF}$, for given $n,m$) that diverge in this limit. From this observation, we pose the central claim of this work: 

\textit{Assuming that the JT gravity SFF in the $\tau$-scaling limit is given by the universal RMT result, the finiteness of this result requires that all terms scaling as $e^{(1+n)S_0}$ mutually cancel.}\footnote{It is conceivable that nonperturbative effects cancel perturbative divergences instead. However, none of the nonperturbative corrections known to the authors are simple powers of the expansion parameter, but rather terms of order e.g. $e^{-e^{S_0}}$. Corrections of this type could not cancel perturbative divergences (which are $\order{e^{-nS_0}}$). This suggests that our prediction still holds, and indeed that the perturbative and nonperturbative parts of the SFF are individually finite. All results we have obtained until now are consistent with this claim.}

Let us elaborate on this claim: In order for different terms of \eqref{eq:SFF_t_expansion} to be able to cancel, they must have the same order in $e^{S_0},$ $\beta$ and $\tau$ \FH{respectively}. \FH{As an example, take again the highest order contribution from above, diverging as $e^{gS_0}$. The terms contributing at this order will all have $n+m=3g-1$, as well as $j+k=n+m$ in the double sum, so as to get the correct power of $t=e^{S_0}\tau$. Indeed, the only nonzero term in the double sum that satisfies this will have $j=m$ and $k=n$. Summing all the terms \TW{in the corresponding contribution in \cref{eq:SFF_t_expansion}}, i.e. summing over partitions of $3g-1$ into two integers $n,m$, we find that the following combination must cancel:}
\begin{equation}
    \FH{\sum_{\substack{n,m=0\\n+m=3g-1}}C^{\WT{(g)}}_{n,m}\frac{\sqrt{\beta^2+t^2}}{\pi}\frac{n!m!2^{n+m-1}}{\gamma^{n+m+1}}\binom{m}{m}\binom{n}{n}i^{n+m}(-1)^m\beta^0t^{m+n}=0}
\end{equation}
\FH{More generally,} it is easy to see that this can only happen for terms with common $n+m$\FH{, and that contributions cannot cancel between different $g$, as this would give extra powers of $e^{S_0}$ for the same power of $\tau$}. Hence, we need to sum up the contributions from all partitions $(n,m)$ of $\FH{n+m}$ for fixed relevant $g$, separately for each relevant\footnote{Relevant in this context means $g>1$ and $j+k>2g$ in the sum \eqref{eq:SFF_t_expansion}. Note also that terms with $j+k<2g$ are suppressed in the $\tau$-scaling limit, while $j+k=2g$ provides precisely the universal part.} power of $\beta$ and $\tau$. Doing so will provide constraint equations that some set of coefficients $C^{\WT{(g)}}_{n,m}$ of the WP volumes need to satisfy.

Consider for example any relevant fixed $g>1$ and $\FH{n+m}$. The first equation we obtain from the above procedure (\FH{upon cancelling global factors and for} $j+k=m+n$) is 
\begin{equation}\label{eq:leading_constraint}
    \sum_{\substack{n,m=0\\n+m\FH{\text{~fixed}}}}C^{\WT{(g)}}_{n,m}n!m!(-1)^m=
    \sum_{\substack{n\geq m\\n+m\FH{\text{~fixed}}}}\FH{\frac{1}{1+\delta_{n,m}}}C^{\WT{(g)}}_{n,m}n!m!\left[(-1)^n+(-1)^m\right]=0,
\end{equation}
\FH{where $\delta_{n,m}$ is the Kronecker delta.} A well-known property of the WP volumes is that the coefficients are symmetric, $C^{\WT{(g)}}_{n,m}=C^{\WT{(g)}}_{m,n}$. We can thus immediately conclude that \eqref{eq:leading_constraint} is always satisfied for odd $n+m$, as then the $(n,m)$ term automatically cancels with $(m,n)$.

However, for even $n+m$, this trivial cancellation will not occur, and \eqref{eq:leading_constraint} is a nontrivial constraint on the WP coefficients. To check this claim for plausibility, let us construct the first example of such a nontrivial cancellation. For $g=2$, the leading equation for maximal $n+m=3g-1=5$ is trivial, while the next-to-leading equations will contribute at order $e^{(5-2g)S_0}=e^{S_0}$, and hence provide the universal part.

The first nontrivial example thus requires $g=3$, and again maximal $n+m=3g-1=8$. Taking the contributions from $j+k=8$ also maximal, we find the following constraint equation:
\begin{equation}\label{eq:g=3}
	280 C^{\WT{(3)}}_{8,0}-35C^{\WT{(3)}}_{7,1}+10C^{\WT{(3)}}_{6,2}-5C^{\WT{(3)}}_{5,3}+2C^{\WT{(3)}}_{4,4}=0
\end{equation}
We can check this by plugging in the coefficients given by \cite{Do2011}
\begin{equation}
\begin{aligned}
	C^{\WT{(3)}}_{8,0}=\frac{1}{856141332480}&,~
    C^{\WT{(3)}}_{7,1}=\frac{1}{21403533312}&,~
    C^{\WT{(3)}}_{6,2}=\frac{77}{152882380800}\\\\
    C^{\WT{(3)}}_{5,3}=\frac{503}{267544166400}&,~
    C^{\WT{(3)}}_{4,4}=\frac{607}{214035333120},
\end{aligned}
\end{equation}
which indeed solve \eqref{eq:g=3}. One can now leave $g$ fixed and vary $n+m$, obtaining a similar equation for a different set of coefficients. In the example at hand however, we would only find the trivial cancellation for $n+m=7$, as well as the result $n+m=6$, which provides the $\beta$-independent part of the universal result. 

However, we could also leave e.g. $n+m=8$ fixed, and simply take terms with $j+k<8$, i.e. contributing at a lower -- but still divergent -- order in $e^{S_0}$. In this way, we can find additional equations for the same set of coefficients as before, taking the form
\begin{align}
    \sum_{\substack{n,m=0\\n+m\FH{\text{~fixed}}}}C^{\WT{(g)}}_{n,m}n!m!(&-1)^m(n-m)=0\label{eq:NL_constraint}\\
    \sum_{\substack{n,m=0\\n+m\FH{\text{~fixed}}}}C^{\WT{(g)}}_{n,m}n!m!(&-1)^m(n-m)^2=0\label{eq:NNL_constraint}\\
    &\vdots\nonumber\\
    \sum_{\substack{n,m=0\\n+m\FH{\text{~fixed}}}}C^{\WT{(g)}}_{n,m}n!m!(&-1)^m(n-m)^l=0\label{eq:Lth_constraint}\\
    &\vdots\nonumber
\end{align}

If the leading order equation is nontrivial, it follows that \eqref{eq:NL_constraint} is trivially satisfied by cancellation of the $(n,m)$ contribution with the $(m,n)$ one, reminiscent of what happens for odd $n+m$ at leading order. However, we can also see that \eqref{eq:NNL_constraint} is a nontrivial constraint equation which is linearly independent from \eqref{eq:leading_constraint}\footnote{In particular, $C_{n,n}$ is not involved in \eqref{eq:NNL_constraint}.}.

For $g=3$, the combination in \eqref{eq:NNL_constraint} provides the $\beta^2$ part of the universal result. Likewise, the analogue of \eqref{eq:NL_constraint} for $n+m=7$ gives the $\beta$-linear part. Generally however, we will obtain a hierarchy of constraints for the coefficients belonging to a given $n+m$, requiring alternately trivial and nontrivial cancellations, until we arrive at a combination that contributes in the universal limit. The general form \eqref{eq:Lth_constraint} of these constraints is readily determined by summing up the contributions to a given $n+m$, yielding
\begin{equation}
    \sum_{\substack{n,m=0\\n+m\FH{\text{~fixed}}}}C^{\WT{(g)}}_{n,m}n!m!\sum_{\substack{j,k=0\\j+k=\FH{n+m}-l}}
    \pmqty{m\\j}\pmqty{n\\k}(-1)^j=0,
\end{equation}
and then cleverly subtracting combinations of the constraints appearing before, i.e. for smaller $l$. For a more rigorous proof, see \cref{app:proof}.

The first example of another constraint than \eqref{eq:leading_constraint} appearing is for $g=4$, where the leading order term vanishes by symmetry. The first (and only) nontrivial constraint for the coefficients with $n+m=3g-1=11$ is given by \eqref{eq:NL_constraint}, which can be simplified to
\begin{align}
    3630 C^{\WT{(4)}}_{11,0}-270C^{\WT{(4)}}_{10,1}+42C^{\WT{(4)}}_{9,2}-10C^{\WT{(4)}}_{8,3}+3C^{\WT{(4)}}_{7,4}-\frac{5}{7}C^{\WT{(4)}}_{6,5}=0.
\end{align}
This equation, as expected, is satisfied by the coefficients obtained by solving the topological recursion,
\begin{equation}
\begin{aligned}
			C^{\WT{(4)}}_{11,0}=\frac{1}{650941377911193600}&,~
			C^{\WT{(4)}}_{10,1}=\frac{1}{8453784128716800}&,~
			C^{\WT{(4)}}_{9,2}=\frac{149}{59176488901017600}\\\\
			C^{\WT{(4)}}_{8,3}=\frac{947}{46026158034124800}&,~
			C^{\WT{(4)}}_{6,5}=\frac{487}{3287582716723200}.
\end{aligned}
\end{equation}
Furthermore, for $g=4$, we find the first constraint on coefficients associated to a non-maximal value of $n+m=3g-2=10$. The constraint is of the type \eqref{eq:leading_constraint} and reads
\begin{align}
    -\frac{5}{56} C^{\WT{(4)}}_{5,5}+\frac{3}{14} C^{\WT{(4)}}_{6,4}-\frac{3}{8} C^{\WT{(4)}}_{7,3}+ C^{\WT{(4)}}_{8,2}-\frac{9}{2} C^{\WT{(4)}}_{9,1}+45C^{\WT{(4)}}_{10,0}=0,
\end{align}
which is indeed satisfied by the relevant coefficients,
\begin{equation}
\begin{aligned}
        C^{\WT{(4)}}_{5,5}=\frac{533\pi^2}{7134511104000}&,~
	C^{\WT{(4)}}_{4,6}=\frac{1081\pi^2}{20547391979520}&,~
	C^{\WT{(4)}}_{3,7}=\frac{16243\pi^2}{898948399104000}\\\\
	C^{\WT{(4)}}_{2,8}=\frac{53\pi^2}{19025362944000}&,~
	C^{\WT{(4)}}_{1,9}=\frac{149\pi^2}{924632639078400}&,~
	C^{\WT{(4)}}_{0,10}=\frac{23\pi^2}{9246326390784000}.
\end{aligned}
\end{equation}

\TW{As a further check, we evaluated the constraints up to $g=5$ and \FH{verified} that they are satisfied (cf. \cref{app:g_5}). We \FH{refrain} from going to even higher orders in checking the constraints and \FH{proceed} to make some general claims about the constraint hierarchies.} 

First, in order for $n+m$ to admit a nontrivial constraint, we must have
\begin{equation}
    2g+1<\FH{n+m}\leq 3g-1.
\end{equation}
Hence, there are in total $3g-1-2g=g-1$ different sets of constrained coefficients, i.e. $g-1$ different constraint hierarchies. In these hierarchies, we will find in principle $n+m-2g$ linearly independent constraints, but every other one of those will be trivially satisfied. This leaves the number of nontrivial constraints for given $n+m$ as 
\begin{equation}
    \begin{aligned}
        &\frac{n+m}{2}-g&&:n+m\mathrm{~even}\\
        &\frac{n+m-1}{2}-g&&:n+m\mathrm{~odd}.
    \end{aligned}
\end{equation}

An interesting question might be how strongly these equations constrain the WP coefficients. The answer, unfortunately, is `not very'. To this end, note that the number of constrained coefficients in a given hierarchy is equivalent to the number of (unordered) pairwise partitions of an integer $n+m$. This number is clearly $\frac{n+m}{2}+1$ for even $\FH{n+m}$ and $\frac{n+m+1}{2}$ for odd $\FH{n+m}$. Hence, we observe that there are 
\begin{equation}
    \#\mathrm{\,variables}-\#\mathrm{\,constraints}=1+g
\end{equation}
coefficients more per hierarchy than can be determined by the constraints. Notably, this number is independent of $n+m$, but it still tells us that there will never be enough equations to actually determine any of the WP coefficients unambiguously, and indeed the problem gets worse as you increase $g$.

\section{Discussion}\label{sec:discussion}
To summarise, we have shown that the Weil-Petersson volumes, which arise as natural objects in the JT gravity path integral, obey some rather nontrivial constraints on their coefficients. These constraints were predicted using nonperturbative, universal information from the dual matrix model. It is this universality that is one of the strengths of our result: we did not need to rely on a particular nonperturbative completion either on the matrix model or on the gravity side. All that was required is the universal limit of the matrix model correlation function we studied, the spectral form factor. 

The striking success of applying RMT universality in the JT/RMT context suggests several further research directions. An obvious and interesting generalisation of our results would be to compute higher $n$-point functions and see if there are similar constraints on the WP volumes appearing therein. However, identifying the correct analogue of the $\tau$-scaling limit that we used for the SFF has proven somewhat difficult so far. A possible resolution would be that the genus expansion for $n>2$ simply cannot capture the universal limit, as it does in the $n=2$ case; a result that would be curious since there does not seem to be anything special about the 2-point function that makes our prediction hold there and not elsewhere. 

Assuming that it is possible to generalise our results to higher $n$, another obvious target would be to try and prove the cancellations directly from the Mirzakhani recursion or (perhaps more simply, and certainly more generally) the topological recursion. As evident from the coefficients cited above, the WP volumes encode information about hyperbolic surfaces in a highly nontrivial manner, and the resulting coefficients need to conspire very precisely to be able to produce a finite result in the $\tau$-scaling limit. It is not far-fetched then to suspect there to be a mechanism \textit{at the level of the recursion} to ensure that the coefficients take values that are compatible with a finite universal limit.

Possibly useful tools to investigate such a mechanism could be provided by intersection theory, which has been used in \cite{BlommaertAndFriends} to identify the same cancellations as we have reported in this work. Some attention has also been directed towards resonance and resurgence of the JT gravity genus expansion \cite{Gregori2021,Baldino2022}, though mainly at the level of the free energy, and it is not immediately clear whether such considerations could be profitably used for the questions at hand.

Finally, since the physical argument that led us to identifying the constraint relations -- the existence of a finite universal limit of the matrix model dual -- isn't particularly fine-tuned to JT gravity (see e.g. \cite{Witten2020}), it seems expedient to try to apply our reasoning to other gravitational models that are often studied in conjunction with JT, for example Liouville quantum gravity or minimal string models. Doing so might help better elucidate the relation of universality and (quantum) gravity if indeed there is any. Another option would be to simply work directly with matrix models defined by the topological recursion, e.g. the one constructed from the spectral density $\tanh\sqrt{E}$. This one is particularly interesting because it seems like the more natural choice for JT gravity when using its description as an $\text{SL}(2,\mathds{R})$ $BF$-theory\footnote{We thank Thomas Mertens for explaining this point to us.}. Here, the metric information is encoded in a flat $\text{SL}(2,\mathds{R})$ connection, for which the natural Plancherel density is the hyperbolic tangent, rather than the hyperbolic sine. However, restricting to smooth geometries (cf. \cite{Mertens2019}), one selects only one component of the connection, exchanging the $\tanh$ for a $\sinh$ in the process (i.e. going over to a description as a $\text{SL}^+(2,\mathds{R})$ $BF$-theory) \cite{Blommaert2019a}. A similar issue appears in the ``quantum particle in $AdS_2$'' description of \cite{Yang2018}, where summing over different $\text{SL}(2,\mathds{R})$ representations, the spectral density tracks a kind of winding number that would correspond to singular geometries in JT gravity, hence requiring regularisation in the form of an infinite imaginary magnetic field in the quantum mechanical system to arrive at the JT result.

\section{Conclusion}
We have here proposed the systematic use of universal RMT results to uncover new relations among functions defined over the moduli space of 2-dimensional manifolds. The essence of this program is the conjectured existence -- well motivated on physical grounds -- of a regime where a strict equivalence between low-dimensional quantum gravity models and the universal correlations given by RMT holds. 

As a first key step in this direction, a particular form of this equivalence has been made explicit in \cite{StanfordAndFriends}  by identifying the regime of parameters where the 2-point functions of Jackiw-Teitelboim gravity are conjectured to be given by the corresponding RMT correlators for the GUE ensemble. Assuming the validity of this conjecture, that includes both perturbative and non-perturbative contributions, we found  hitherto unknown identities among the numerical factors of the polynomial Weil-Petersson volumes on the JT side. 

Since the non-trivial correctness of the identities for $n=2$ is now firmly established and checked against exact results, the possible extension of this program for higher order functions (and the corresponding Weil-Petersson volumes) as well as to other types of models as discussed  here is a promising route to merge RMT with the theory of hyperbolic manifolds.

\section*{Acknowledgements}
\label{sec:acknowledgments}

We thank T. Mertens for illuminating discussions, and A. Blommaert, D. Stanford, J. Kruthoff and S. Yao for very helpful conversations and for sharing some of their unpublished results during preparation of \cite{BlommaertAndFriends, StanfordAndFriends}.

We acknowledge financial support from the Deutsche Forschungsgemeinschaft (German Research Foundation) through Ri681/15-1 within the Reinhart-Koselleck Programme.
\newpage
\appendix

\section{Proof of the form of the constraint hierarchies}\label{app:proof}
As stated above, for a chosen value of genus $g$ and $n+m$ one finds a hierarchy of constraints\FH{. The $l$th constraint is} given by 
\begin{align}
    \FH{K}_l(\FH{n+m})&\coloneqq\sum_{\substack{n,m=0\\n+m\FH{\text{~fixed}}}}C^{\WT{(g)}}_{n,m}n!m!\sum_{\substack{j,k=0\\j+k=\FH{n+m}-l}}
    \pmqty{m\\j}\pmqty{n\\k}(-1)^j\\
    &\eqqcolon\sum_{\substack{n,m=0\\n+m\FH{\text{~fixed}}}}C^{\WT{(g)}}_{n,m}n!m!P_l(n,m),\label{app:defPl}
\end{align}
for values of $l$ such that $j+k>2g$. We proceed to prove that a linearly independent combination of these constraints takes the form
\begin{equation}
    \FH{K}_l^{\FH{\text{simplified}}}(\FH{n+m}):=\sum_{\substack{n,m=0\\n+m\FH{\text{~fixed}}}}C^{\WT{(g)}}_{n,m}n!m!\qty(-1)^m \qty(n-m)^l. \label{app:constraint}
\end{equation}
As a first step by using the property $\binom{n}{m}=\binom{n}{n-m}$ of the binomial coefficient it is useful to rewrite the $P_l(n,m)$ into the more convenient form
\begin{align}
    P_l(n,m)&=\sum_{\substack{j,k=0\\j+k=\FH{n+m}-l}}
    \binom{m}{m-j}\binom{n}{n-k}(-1)^j\\
    &=\qty(-1)^m\sum_{\substack{\delta,\gamma=0\\\delta+\gamma=l}}(-1)^{\gamma}
    \binom{m}{\gamma}\binom{n}{\delta},
\end{align}
where $\gamma=m-j$ and $\delta=n-k$.\\
Next, we show that $P_l(n,m)$ is a polynomial of degree $l$ in m, as well as in n. To do so, it suffices to show that each individual term in $P_l(n,m)$ is a polynomial. For $a<b$, it holds that
\begin{align}
    \binom{b}{a}=\frac{b!}{a!\qty(b-a)!}=\frac{1}{n!}\prod_{j=0}^{a-1}\qty(b-j),
\end{align}
which is a polynomial of degree $a$ in $b$. Hence, both binomial coefficients appearing in $P_l(n,m)$ are polynomials in $n$ or $m$, and so is their product. Evidently, the degree of these polynomials is $l$, whence
\begin{equation}
    \begin{aligned}
    P_l(n,m)&=\qty(-1)^m\qty[\binom{m}{0}\binom{n}{l}+\qty(-1)^l\binom{m}{l}\binom{n}{0}+\qty(\text{lower order in $n,m$})]\\
    &=\qty(-1)^m\qty[\frac{1}{l!}n^l+\frac{1}{l!}\qty(-m)^l+\qty(\text{lower order in $n,m$})]\\
    &=\qty(-1)^m\qty[\frac{1}{l!}\qty(n-m)^l+\qty(\text{lower order in $n,m$})].
\end{aligned}
\end{equation}
From here, \eqref{app:constraint} follows immediately upon realising that the lower order terms, plugged into \eqref{app:defPl}, all reduce to linear combinations of constraints for smaller $l$ (notice that $n+m$ is a global constant for each term and can be pulled out of the sum). If the previous constraints are satisfied, they can be safely subtracted, leaving the desired expression \eqref{app:constraint}.

\section{Constraints for g=5}\label{app:g_5}
\TW{
\FH{$g=5$ marks the first instance of} multiple constraints on one particular set of coefficients. Using the formulae stated above one finds for $n+m=14$
\begin{align}
    l=&0:\quad 48048 C^{\WT{(5)}}_{14, 0} - 3432 C^{\WT{(5)}}_{13, 1} + 528 C^{\WT{(5)}}_{12, 2} - 132 C^{\WT{(5)}}_{11, 3} + 48 C^{\WT{(5)}}_{10, 4} - 24 C^{\WT{(5)}}_{9, 5} + 16 C^{\WT{(5)}}_{8, 6} - 7 C^{\WT{(5)}}_{7, 7}&&=0,\\
    l=&1:\quad 147147 C^{\WT{(5)}}_{14, 0} - 7722 C^{\WT{(5)}}_{13, 1} + 825 C^{\WT{(5)}}_{12, 2} - 132 C^{\WT{(5)}}_{11, 3} + 27 C^{\WT{(5)}}_{10, 4} - 6 C^{\WT{(5)}}_{9, 5} +  C^{\WT{(5)}}_{8, 6}&&=0.
    \end{align}
For $n+m=13$:
\begin{align}
    l=0: \quad 22308 C^{\WT{(5)}}_{13, 0} - 1452 C^{\WT{(5)}}_{12, 1} + 198 C^{\WT{(5)}}_{11, 2} - 42 C^{\WT{(5)}}_{10, 3} + 12 C^{\WT{(5)}}_{9, 4} - 4 C^{\WT{(5)}}_{8, 5} +  C^{\WT{(5)}}_{7, 6}=0.
\end{align}
Evaluating the coefficients of the WP volume $V_{5,2}$ by the procedure outlined in \cref{app:top_recursion}, one finds
\begin{equation}
\begin{aligned}
			C^{\WT{(5)}}_{14,0}=\frac{1}{1364789730583724949504000}&,~
			C^{\WT{(5)}}_{13,1}=\frac{1}{10831664528442261504000}\\\\
			C^{\WT{(5)}}_{12,2}=\frac{7}{2142527049581985792000}&,~
			C^{\WT{(5)}}_{11,3}=\frac{307}{6561489089344831488000}\\\\
			C^{\WT{(5)}}_{10,4}=\frac{29}{88370223425519616000}&,~
                C^{\WT{(5)}}_{9,5}=\frac{2351}{18747111683842}\\\\
			C^{\WT{(5)}}_{8,6}=\frac{2291}{833204963726327808000}&,~
                C^{\WT{(5)}}_{7,7}=\frac{173}{48603622884035788800},
\end{aligned}
\end{equation}
for the leading order coefficients and
\begin{equation}
\begin{aligned}
			C^{\WT{(5)}}_{13,0}=\frac{29\pi^2}{12185622594497544192000}&,~
			C^{\WT{(5)}}_{12,1}=\frac{7\pi^2}{26781588119774822400}\\\\
			C^{\WT{(5)}}_{11,2}=\frac{257\pi^2}{32547068895559680000}&,~
			C^{\WT{(5)}}_{10,3}=\frac{35521\pi^2}{372811880076410880000}\\\\
			C^{\WT{(5)}}_{9,4}=\frac{11827\pi^2}{21303536004366336000}&,~
                C^{\WT{(5)}}_{8,5}=\frac{12491\pi^2}{7232681976791040000}\\\\
			C^{\WT{(5)}}_{7,6}=\frac{195983\pi^2}{65094137791119360000}&,~
\end{aligned}
\end{equation}
for the next-to-leading order ones. As one can check by plugging those coefficients \FH{in} the constraints given above, they are indeed satisfied.
}
\section{The topological recursion for the JT matrix model}\label{app:top_recursion}

To give a complete picture of the matrix model of \cite{Saad2019}, the objective of this appendix is to \FH{introduce} the topological recursion that is used to compute the perturbative expansion of observables in this matrix model, \FH{using} only the leading genus density of states and the symmetry class of the model as input. \WT{As this method here is primarily used as a tool to compute the WP volumes, we aim at giving a brief minimal-technical presentation, referring for the details to the literature.}

Restricting again to the unitary symmetry class\FH{, we specialise to} the leading order density of states given \FH{by} \cref{eq:rho_0_t}
\begin{align}
    \rho^{\text{JT}}_0(E)=\frac{\gamma}{2\pi^2}\sinh(2\pi\sqrt{2\gamma E}),
\end{align}
where we introduce the superscript to distinguish this specific $\rho_0$ from the general expressions appearing in the following discussion.\\
To formulate the topological recursion, it is best to define the \textit{spectral curve} $y(E)$\WT{\footnote{\WT{Following the notational convention of \cite{Saad2019} which we adopt in the following. The ``spectral curve'' in the sense of e.g. \cite{Eynard2018}, being the mathematically precise definition of this object, following the notation there is given by $\Sigma=\mathbb{P}^1$, with $\mathbb{P}^1$ denoting the Riemann sphere, due to the matrix model being a one-cut matrix model, $x(z)=z^2(=-E)$, $y(z)=y^{\text{JT}}(z)$, $B=\omega_{0,2}$ with the not yet defined objects defined below.}}} via 
\begin{align}\label{eq:def_y_pyhsSheet}
    y(E)=-i\pi \rho_0(E).
\end{align}
Thus, for the case of the JT density of states one finds
\begin{equation}\label{eq:y_JT_x}
    \begin{aligned}
        y^{\text{JT}}(E)&=-i\pi\rho^{\text{JT}}_0(E)=\frac{-i\gamma}{2\pi}\sinh(2\pi\sqrt{2\gamma E})\\
        &=\frac{\gamma}{2\pi}\sin\left(2\pi\sqrt{2\gamma} \qty(-i\sqrt{E})\right).
    \end{aligned}
\end{equation}
This object naturally appears in the context of the \textit{loop equations} for matrix models defined via a potential $V(H)$ (often assumed to be a polynomial), i.e. via the partition function
\begin{align}
    \mathcal{Z}= \int_{\mathbb{E}} e^{-\tr(V(H))}\dd H,
\end{align}
where $\mathbb{E}$ denotes the space of matrices of the symmetry class one wishes to study \cite{Eynard2018,Stanford2019}. \WT{As mentioned in the introduction, this integral is taken to be a ``formal'' integral, meaning that, as one is accustomed to in QFT, the quadratic term in the potential yields a propagator and higher terms of the potential are treated by a formal Taylor expansion of the potential, which can be evaluated by computing Gaussian integrals, i.e. diagrammatically \cite{Eynard2018}. More precisely, the formal matrix model leading to $y^{\text{JT}}$ is the Kontsevich matrix model \cite{Kontsevich1992} for a particular choice of Kontsevich times found by the Taylor expansion of $y^{\text{JT}}$ \cite{Eynard2007c}. The applicability of the topological recursion to this particular case was shown in \cite{Eynard2007b}, thus justifying its use for the spectral curve considered here.}\\
It turns out that for the topological recursion the most important property of the spectral curve is its behaviour at its branch points. Thus it is useful to study this in more detail for the case of $y^{\text{JT}}$. From the first line of \cref{eq:y_JT_x}, one can see quite clearly that the spectral curve's sole branch point is at the origin coming from $\sqrt{E}$. At this point, it is useful to introduce a new coordinate that takes care of the branch-cut structure intrinsically. This coordinate, following the convention of \cite{Saad2019}, is defined via
\begin{align}
    z^2=-E.\label{eq:defZ}
\end{align}
\FH{The branch cut structure is implemented by solving for $z=\pm i\sqrt{E}$ and choosing the negative sign} on the physical sheet (where \cref{eq:def_y_pyhsSheet} holds) and \FH{the positive sign} on the other sheet%\footnote{A generalization of \cref{eq:def_y_pyhsSheet} which indeed is the expression obtained when deriving $y$ from the loop equations, which has been omitted here for brevity, would be $\lim\limits_{\epsilon\rightarrow 0}y(E\pm i \epsilon)=\mp i \pi \rho_0(E)$. }
. Thus, by choosing $z$ as coordinate the branch cut structure is implemented \FH{automatically} and all functions become single valued. As a first example, \FH{consider} the second line of \FH{\eqref{eq:y_JT_x}} to obtain
\begin{align}
	y^{\text{JT}}(z)=\frac{\gamma}{2\pi}\sin(2\pi\sqrt{2\gamma}z),
\end{align}
which is \FH{indeed} single-valued.\\
\FH{I}t is furthermore important to define the \textit{resolvent} via\footnote{\TW{They are not to be \FH{mistaken for} the $R_n(E_1,\dots,E_n)$ considered in the main text.}}
\begin{align}
    R(E)\coloneqq \ev{ \tr(\frac{1}{E-H})}\label{eq:DefRes}.
\end{align}
Analogously to $\rho$ and $Z(\beta)$, one can of course also consider (connected) correlation functions of resolvents. \FH{Correlators of partition functions can furthermore be computed from the resolvents} by utilizing the generalisation of 
\begin{equation}
    R(E)=-\int_{0}^{\infty}\dd \beta e^{\beta E} Z(\beta) \label{eq:R(Z)},
\end{equation}
to products of resolvents inside the matrix-model expectation value. \FH{Hence after double-scaling, by linearity of the integral,} there is an expansion in $e^{-S_0}$ of the correlation functions of resolvents as 
\begin{align}
    \ev{R(E_1,\dots,E_n)}=\sum_{g=0}^{\infty} \frac{R_{g,n}(E_1,\dots,E_n)}{\qty(e^{S_0})^{2g+n-2}}.
\end{align}
Having defined this object, \FH{we cite the topological recursion} from \cite{Eynard2018} and appl\FH{y it} to the present case. The recursion is most naturally formulated using differential forms and thus one defines\footnote{\TW{It is not notational sloppiness leading to the missing of wedge-product signs but $\omega_{g,n}$ is to be understood as an element of the tensor product of $n$ one-forms.}}
\begin{align}\label{eq:top_rec_diff_forms}
    \omega_{g,n}(z_1,\dots,z_n)=R_{g,n}(E_1,\dots,E_n)\dd E_1 \dots\dd E_n+\delta_{g,0}\delta_{n,2}\frac{\dd E_1\dd E_2}{(E_1-E_2)^2},
\end{align}
where the coordinates $E_i$ are the coordinates on the branched surface interpreted as functions of the ``double-cover'' coordinates $z_i$ defined in \cref{eq:defZ}. Using this definition, one has $\dd E = - 2 z \dd z$ and thus one can write for $(g,n)\neq(0,2)$
\begin{equation}
\begin{aligned}
    \omega_{g,n}=&\qty(-2)^n z_1\dots z_n R_{g,n}(-z_1^2,\dots,-z_n^2) \dd z_1\dots\dd z_n\\
\coloneqq &W_{g,n}(z_1,\dots,z_n) \dd z_1\dots\dd z_n.\label{eq:defW}
\end{aligned}
\end{equation}
For the excluded case, $(g,n)=(0,2)$, one defines $W_{0,2}$ analogously including the additional summand \FH{in \cref{eq:top_rec_diff_forms}}\footnote{\TW{Adding the additional term to $\omega_{0,2}$ seems unreasonable at first sight but as one can see e.g. from the well known relation of $\ev{\rho(E_1)\dots\rho(E_n)}$ to $\ev{R(E_1),\dots,R(E_n)}$, only the discontinuity of $R$ through the cut matters and the additional term does not contribute to this. Thus, it is valid to use $W_{0,2}$ to compute e.g. the corresponding contribution to $\ev{Z(\beta_1) Z(\beta_2)}$. However, one should keep in mind that the added term actually introduces a singularity as $E_1\to E_2$ which was not present in the original $R_{0,2}$.}}. \FH{We proceed to stating} the topological recursion determining the $\omega_{g,n}$ for a spectral curve $y$:\\
First, by definition\footnote{\TW{The definition of $\omega_{0,1}$ is modified by a sign compared to \cite{Eynard2018} to follow the sign convention of \cite{Saad2019}. Using the general result for $R_{0,2}$ for one-cut matrix models, found e.g. in \cite{Stanford2019} one can verify directly that $\omega_{0,2}$ is indeed of this form.}}
\begin{align}
	\omega_{0,1}&=-y\dd E=2zy(z)dz,\\
	\omega_{0,2}&=\frac{\dd z_1 \dd z_2}{\qty(z_1-z_2)^2},
\end{align}
which serves as the input to the recursion. Now, for each branch point $a$ of the spectral curve one defines 
\begin{align}
		K_a(z_1,z)\coloneqq \frac 12 \frac{\int_{\sigma_a(z)}^{z}\omega_{0,2}(z_1,\cdot)}{\omega_{0,1}(z)-\omega_{0,1}(\sigma_a(z))},
\end{align}
\FH{where} $\sigma_a(z)$ \FH{is} a so-called local Galois involution which denotes a function locally exchanging the two sheets meeting at the branch point\footnote{\TW{Put plainly, it is a function mapping the coordinate $z$ to the other coordinate corresponding to the same $E$. For the coordinate defined in \cref{eq:defZ}, $\sigma_0(z)=-z$.\label{fn:galois}}}. Furthermore, the notation $\int \omega(z_1,\cdot)$ with $\omega\propto \dd z_1\dd z_2$ denotes the integration of the one-form whose position is replaced by the $\cdot$. Having defined this, one can write the topological recursion for the unitary symmetry class determining the other $\omega_{g,n}$, using the abbreviating notation of $J$ for the (ordered) set $\qty{z_2,\dots,z_n}$, as
\begin{equation}
    \begin{aligned}
	\omega_{g,n}(z_1,J)=\sum_{a}\underset{z=a}{\Res}K_a(z_1,z) [\omega_{g-1,n+1}(z,\sigma_a(z),J)\\+\underset{\substack{h+h'=g\\ I\cup I'=J}}{\sum^{'}} \omega_{h,1+\abs{I}}(z,I)\omega_{h',1+\abs{I'}}(\sigma_a(z),I')]\label{eq:TopRecursionGen},
    \end{aligned}
\end{equation}
with $\sum^{'}$ denoting the sum excluding the cases $(h,I)=(0,\emptyset)$ and $(h,I)=(g,J)$.\\
To apply this result, taken from \cite{Eynard2018}, it remains to compute $K_a$ for the branch points of the spectral curve one wishes to study. For the spectral curve of present interest, $y^{\text{JT}}$, there is only one branch point $a=0$ and thus only $K_0$ has to be computed. As noted \FH{in \cref{fn:galois}}, $\sigma_0(z)=-z$ and thus 
\begin{equation}
    \begin{aligned}
	K_0(z_1,z)&=\frac 12 \frac{\int_{-z}^{z}\omega_{0,2}(z_1,\cdot)}{\omega_{0,1}(z)-\omega_{0,1}(-z)}=\frac{\frac 12 \dd z_1 \int_{-z}^{z}\frac{\dd z_2}{\qty(z_1-z_2)^2}}{2 z \qty(y(z)-y(-z))\dd z}\\
	&=\frac{\frac 12 \dd z_1 \qty[\frac{1}{z_1-z}-\frac{1}{z_1+z}]}{2 z \qty(y(z)-y(-z))\dd z}\\
        &=\frac{\dd z_1}{z_1^2-z^2}\frac{1}{2\qty(y(z)-y(-z))\dd z}.
    \end{aligned}
\end{equation}
Using the antisymmetry of $y^{\text{JT}}$ one has
\begin{align}
    K_0(z_1,z)=\frac{1}{z_1^2-z^2} \frac{1}{4 y(z)}\frac{\dd z_1}{\dd z}.
\end{align}
Putting this into \cref{eq:TopRecursionGen} one can, after noting that on the RHS one of the $\dd z$ is cancelled by the denominator of $K_0$, drop the differentials and write the equation in terms of the $W_{g,n}$, yielding
\begin{equation}
    \begin{aligned}
	W_{g,n}(z_1,J)=\underset{z=0}{\Res}\bigg\{\frac{\dd z}{z_1^2-z^2} \frac{1}{4 y(z)} \Big [ W_{g-1,n+1}(z,-z,J)\\+\underset{\substack{h+h'=g\\ I\cup I'=J}}{\sum^{'}} W_{h,1+\abs{I}}(z,I)W_{h',1+\abs{I'}}(-z,I')\Big ]\bigg\},\label{eq:TopRec}
    \end{aligned}
\end{equation}
in accordance with the relation given in \cite{Saad2019}. Note, that this formula only holds for the unitary symmetry class, \FH{and that} were one to consider a different symmetry class, \FH{but} the same spectral curve, additional difficulties appear.\\
\FH{After arriving at} the topological recursion, a useful final step is to relate them to the WP volumes, as \FH{those} are the objects of interest for the main text. \FH{T}he definition of the $W_{g,n}$ in \cref{eq:defW} and the relation of the correlators of resolvents to those of  partition functions (cf. \cref{eq:R(Z)}) imply
\begin{align}\label{eq:lap_trafo}
	W_{g,n}(z_1,\dots,z_n)=\prod_{i=1}^{n}\qty[2z\int_{0}^{\infty}\dd{\beta_i}e^{-\beta_i z_i^2}]Z_{g,n}(\beta_1,\dots,\beta_n).
\end{align}
Thus, one can compute $W_{g,n}^\text{JT}$, the object derived from the correlators of partition functions in JT gravity, by \FH{plugging} the general expression for $Z_{g,n}$, given in \cref{eq:genus_expansion}, into \eqref{eq:lap_trafo}\footnote{\TW{The cases of $(g,n)=(0,1)$, $(g,n)=(0,2)$ where this expression does not hold are exactly the cases where the $W_{g,n}$ are given as input to the topological recursion, \FH{hence} already known.}}. This yields
\begin{equation}
    \begin{aligned}
		W_{g,n}^\text{JT}(z_1,\dots,z_n)%&=\prod_{i=1}^{n}\qty[2z_i\int_{0}^{\infty}\dd{\beta_i}e^{-\beta_i z_i^2}\int_{0}^{\infty}b_i \dd b_i Z^{t}(\beta_i,b_i)]V_{g,n}(b_1,\dots,b_n)\\
		&=\prod_{i=1}^{n}\qty[\int_{0}^{\infty}b_i \dd b_i 2z_i\int_{0}^{\infty}\dd{\beta_i}e^{-\beta_i z_i^2} Z^{t}(\beta_i,b_i)]V_{g,n}(b_1,\dots,b_n).\label{eq:W(V)_interm}
    \end{aligned}
\end{equation}
Now one can perform the Laplace transform of the trumpet using \cite{Gradshteyn2000}[\FH{formula} 17.13.30] and use this result to find
% \begin{equation}
% 	\begin{aligned}
% 	\int_{0}^{\infty}\dd{\beta_i}e^{-\beta_i z_i^2} Z^{t}(\beta_i,b_i)&=\int_{0}^{\infty}\dd{\beta_i}\sqrt{\frac{\gamma}{2\pi}}e^{-\beta_i z_i^2}\frac{e^{-\frac{\gamma b_i^2}{2\beta_i}}}{\sqrt{\beta_i}}=\sqrt{\frac{\gamma}{2\pi}}\sqrt{\pi}\frac{e^{\sqrt{-2b_i^2 \gamma z_i^2}}}{z_i}\\
% 			&=\frac{\sqrt{2\gamma}}{2z_i}e^{-\sqrt{2\gamma}b_i z_i}.
% 	\end{aligned}
% \end{equation}
% Plugging this into \cref{eq:W(V)_interm}, one obtains
\begin{align}
    W_{g,n}^\text{JT}(z_1,\dots,z_n)=\prod_{i=1}^{n}\qty[\sqrt{2\gamma}\int_{0}^{\infty}b_i \dd b_i e^{-\sqrt{2\gamma}b_i z_i}]V_{g,n}(b_1,\dots,b_n).\label{eq:W(V)}
\end{align}
Thus, for JT gravity, $W_{g,n}$ is the (modified) Laplace transform of the Weil-Petersson volume of genus $g$ and $n$ geodesic boundaries multiplied with \FH{by} boundary lengths. To facilitate an easier comparison, it is useful to observe that if $W_{g,n}(z_1,\dots,z_n)$ are the \FH{outputs} of the topological recursion in \cref{eq:TopRec} for the spectral curve $y(z)$, then $W'(z_1,\dots,z_n)\coloneqq \sqrt{2\gamma}^n W(\sqrt{2\gamma}z_1,\dots,\sqrt{2\gamma}z_n)$ are the \FH{outputs} of the topological recursion for the spectral curve $y'(z)\coloneqq 2\gamma y(\sqrt{2\gamma}z)$ as one can check simply by putting the primed objects into the topological recursion (\cref{eq:TopRec}) \cite{Saad2019}. Thus, one can set w.l.o.g $\gamma=\frac 12$, in which case the relation between the $W_{g,n}^\text{JT}$ and the WP volumes becomes the usual Laplace transform.\\
As the general form of the WP volumes is known from \cite{Mirzakhani2007}, it is convenient to compute the general version of the RHS of \cref{eq:W(V)} to obtain a simple relation between coefficients appearing in the $W_{g,n}^{\text{JT}}$ and those in the WP volumes. The general form of the WP volumes (from which \cref{eq:def_WP} is derived) is given by 
\begin{align}
	V_{g,n}(\vec{L})=\sum_{\vec{a}}^{\norm{a}_1\leq 3g-3+n}C^{\vec{a}}_{g,n} \prod_{i=1}^{n}L_i^{2a_i},
\end{align}
with $\vec{L}\in \mathbb{R}_+^n$, $\vec{a}\in \mathbb{N}_0^n$, $\norm{a}_1=\sum_{i=1}^n a_i$ and positive constants $C^{\vec{a}}_{g,n}\in \pi^{6g-g+2n-2\norm{a}_1} \cdot \mathbb{Q}$ \footnote{\TW{The general version of the WP coefficients needed here is related to the ones for $n=2$ considered in the main text via $C_{g,2}^{\qty(n,m)}=C^{(g)}_{n,m}$}}.
The WP volumes are symmetric polynomials in the boundary lengths and so the coefficients $C^{\vec{a}}_{g,n}$ are symmetric under all permutations of entries in $\vec{a}$. Thus, one can compute the $W_{g,n}^\text{JT}$ by performing the Laplace transform in \cref{eq:W(V)}, using \cite{Gradshteyn2000}[\FH{formula} 17.13.2] as
\begin{align}
	W_{g,n}^\text{JT}(z_1,\dots,z_n)=\sum_{\vec{a}}C^{\vec{a}}_{g,n} \prod_{i=1}^{n}\frac{\qty(2a_i+1)!}{\qty(z_i^2)^{a_i+1}}\label{eq:W(C)}.
\end{align}
This enables one to read off the coefficients of the WP volumes directly from the result of the topological recursion, \FH{since by invoking the duality of JT gravity and the matrix model \cite{Saad2019} it holds that the $W_{g,n}$ found by performing the topological recursion using $y^{\text{JT}}$ equal the $W^{\text{JT}}_{g,n}$} \footnote{\TW{Actually, at this stage one can see this explicitly as by a result of Eynard and Orantin \cite{Eynard2007c}, the RHS of \cref{eq:W(V)} (for $\gamma=\frac 12$) holds for the $W_{g,n}$ obtained by using $y^{\text{JT}}$.}}.\\
This way of computing the WP volumes \FH{is} the way that was used in this work \FH{and in \cite{Miedaner2023}} to obtain the WP volumes that could not be found in the literature.

\section{Collection of relevant WP volumes}\label{app:WP}
\TW{
For the convenience of the reader, we include here a collection of the relevant WP volumes for $n=2$, obtained by the method stated in \cref{app:top_recursion}. To abbreviate the presentation of the volumes, we introduce the notation \FH{of} \cite{Do2011}
\begin{align}
    m_{a_1,\cdots, a_n}(L_1,\cdots,L_n):=\sum_{\pi\in S(n)}\prod_{i=1}^{n}L_i^{2 a_{\pi(i)}},
\end{align}
where $S(n)$ denotes the permutation group for a set of order $n$. This notation is introduced to put the symmetry of the WP volumes into the polynomial part so that each of the coefficients only appears once. The volumes we give here up to $g=3$ coincide with the ones published in \cite{Do2011}, \FH{the higher ones were computed in \cite{Miedaner2023}}.
\begin{align}
    &\begin{aligned}
        V_{1,2}(L_1,L_2)&=\frac{1}{192}m_{2}+\frac{1}{96}m_{1,1}+\frac{\pi^2}{12} m_{1}+\frac{\pi^4}{4}
    \end{aligned}
    \\
    &\begin{aligned}
        V_{2,2}(L_1,L_2)=&\frac{1}{4423680}m_{5}+\frac{1}{294912}m_{4,1}+\frac{29}{2211840}m_{3,2}\\
            		&+\frac{11\pi^2}{276480}m_{4}+\frac{29\pi^2}{69120}m_{3,1}+\frac{7\pi^2}{7680}m_{2,2}\\
                        &+\frac{19\pi^4}{7680}m_{3}+\frac{181\pi^4}{11520}m_{2,1}\\
                        &+\frac{551\pi^2}{8640}m_{2}+\frac{7\pi^6}{36}m_{1,1}\\
                        &+\frac{1085\pi^8}{1728}m_{1}+\frac{787\pi^{10}}{480}
    \end{aligned}
    \\
    &\begin{aligned}
		 	V_{3,2}(L_1,L_2)=&\frac{1}{856141332480}m_{8}+\frac{1}{21403533312}m_{7,1}+\frac{77}         {152882380800}m_{6,2}\\
                                &+\frac{503}{267544166400}m_{5,3}+\frac{607}{214035333120}m_{4,4}\\
                                &+\frac{17\pi^2}{22295347200}m_{7}+\frac{77\pi^2}{3185049600}m_{6,1}+\frac{17\pi^2}{88473600}m_{5,2}+\frac{1121\pi^2}{2229534720}m_{4,3}\\
                                &+\frac{1499\pi^4}{7431782400}m_{6}+\frac{899\pi^4}{185794560}m_{5,1}+\frac{10009\pi^4}{371589120}m_{4,2}+\frac{191\pi^4}{4128768}m_{3,3}\\
                                &+\frac{3859\pi^6}{139345920}m_{5}+\frac{33053\pi^6}{69672960}m_{4,1}+\frac{120191\pi^6}{69672960}m_{3,2}\\
                                &+\frac{195697\pi^8}{92897280}m_{4}+\frac{110903\pi^8}{4644864}m_{3,1}+\frac{6977\pi^8}{138240}m_{2,2}+\\
                                &+\frac{37817\pi^{10}}{430080}m_{3}+\frac{2428117\pi^{10}}{4147200}m_{2,1}+\\
                                &+\frac{5803333\pi^{12}}{3110400}m_{2}+\frac{18444319\pi^{12}}{3110400}m_{1,1}\\
                              &+\frac{20444023\pi^{14}}{1209600}m_{1}+\frac{2800144027\pi^{16}}{65318400}	
		 \end{aligned}
\end{align}
\begin{equation}
\begin{aligned}
        V_{4,2}(L_1,L_2)=&\frac{1}{650941377911193600}m_{11} + \frac{1}{8453784128716800}m_{10,1} + \frac{149}{59176488901017600}m_{9,2}\\
        &+ \frac{947}{46026158034124800}m_{8,3} + \frac{1781}{23013079017062400}m_{7,4} + \frac{487}{3287582716723200}m_{6,5}\\
        &+ \frac{23 \pi^{2}}{9246326390784000}m_{10} + \frac{149 \pi^{2}}{924632639078400}m_{9,1} + \frac{53 \pi^{2}}{19025362944000}m_{8,2}\\
        &+ \frac{16243 \pi^{2}}{898948399104000}m_{7,3} + \frac{1081 \pi^{2}}{20547391979520}m_{6,4} + \frac{533 \pi^{2}}{7134511104000}m_{5,5}\\
        &+ \frac{18691 \pi^{4}}{10787380789248000}m_9 + \frac{55367 \pi^{4}}{599298932736000}m_{8,1} + \frac{189851 \pi^{4}}{149824733184000}m_{7,2}\\
        &+ \frac{38789 \pi^{4}}{6115295232000}m_{6,3} + \frac{41987 \pi^{4}}{3057647616000}m_{5,4}\\
        &+ \frac{43591 \pi^{6}}{64210599936000}m_8 + \frac{407821 \pi^{6}}{14046068736000}m_{7,1} + \frac{4927249 \pi^{6}}{16052649984000}m_{6,2}\\
        &+ \frac{654223 \pi^{6}}{573308928000}m_{5,3} + \frac{39947 \pi^{6}}{22932357120}m_{4,4}\\
        &+ \frac{105541 \pi^{8}}{642105999360}m_{7} + \frac{17546603 \pi^{8}}{3210529996800}m_{6,1} + \frac{212383 \pi^{8}}{4954521600}m_{5,2} + \frac{25760323 \pi^{8}}{229323571200}m_{4,3}\\
        &+ \frac{13624007 \pi^{10}}{535088332800}m_6 + \frac{42308743 \pi^{10}}{66886041600}m_{5,1} + \frac{467464817 \pi^{10}}{133772083200}m_{4,2} + \frac{19096223 \pi^{10}}{3185049600}m_{3,3}\\
        &+ \frac{634238489 \pi^{12}}{250822656000}m_5 + \frac{4465379809 \pi^{12}}{100329062400}m_{4,1} + \frac{5043377833 \pi^{12}}{31352832000}m_{3,2}\\
        &+ \frac{275930395973 \pi^{14}}{1755758592000}m_4 + \frac{798137682887 \pi^{14}}{438939648000}m_{3,1} + \frac{2223110269 \pi^{14}}{580608000}m_{2,2}\\
        &+ \frac{15253048628171 \pi^{16}}{2633637888000}m_3 + \frac{2743831363 \pi^{16}}{69984000}m_{2,1}\\
        &+ \frac{5652202930679 \pi^{18}}{49380710400}m_2 + \frac{18239563926361 \pi^{18}}{49380710400}m_{1,1}\\
        &+ \frac{70726245137 \pi^{20}}{70543872}m_1 + \frac{909612310986473 \pi^{22}}{362125209600}
\end{aligned}
\end{equation}
    %\chapter{Temp}
\begin{equation}
    \begin{aligned}
        V_{5,2}=&\frac{1}{1364789730583724949504000}m_{14} + \frac{1}{10831664528442261504000}m_{13,1} + \frac{7}{2142527049581985792000}m_{12,2}\\
        &+ \frac{307}{6561489089344831488000}m_{11,3} + \frac{29}{88370223425519616000}m_{10,4} + \frac{2351}{1874711168384237568000}m_{9,5}\\
        &+ \frac{2291}{833204963726327808000}m_{8,6} + \frac{173}{48603622884035788800}m_{7,7}\\
        &+ \frac{29 \pi^{2}}{12185622594497544192000}m_{13} + \frac{7 \pi^{2}}{26781588119774822400}m_{12,1} + \frac{257 \pi^{2}}{32547068895559680000}m_{11,2}\\
        &+ \frac{35521 \pi^{2}}{372811880076410880000}m_{10,3} + \frac{11827 \pi^{2}}{21303536004366336000}m_{9,4} + \frac{12491 \pi^{2}}{7232681976791040000}m_{8,5}\\
        &+ \frac{195983 \pi^{2}}{65094137791119360000}m_{7,6}\\
        &+ \frac{12533 \pi^{4}}{3645271716302684160000}m_{12} + \frac{111059 \pi^{4}}{341744223403376640000}m_{11,1} + \frac{513881 \pi^{4}}{62135313346068480000}m_{10,2}\\
        &+ \frac{342973 \pi^{4}}{4142354223071232000}m_{9,3} + \frac{155051 \pi^{4}}{394509926006784000}m_{8,4} + \frac{3691217 \pi^{4}}{3797158037815296000}m_{7,5}\\
        &+ \frac{28393333 \pi^{4}}{21698045930373120000}m_{6,6}\\
        &+ \frac{746359 \pi^{6}}{256308167552532480000}m_{11} + \frac{906701 \pi^{6}}{3883457084129280000}m_{10,1} + \frac{3280321 \pi^{6}}{665735500136448000}m_{9,2}\\
        &+ \frac{20793847 \pi^{6}}{517794277883904000}m_{8,3} + \frac{934747 \pi^{6}}{6164217593856000}m_{7,4} + \frac{23560181 \pi^{6}}{81367672238899200}m_{6,5}\\
        &+ \frac{822676861 \pi^{8}}{512616335105064960000}m_{10} + \frac{5502349099 \pi^{8}}{51261633510506496000}m_{9,1} + \frac{291625027 \pi^{8}}{158214918242304000}m_{8,2}\\
        &+ \frac{680389207 \pi^{8}}{56957370567229440}m_{7,3} + \frac{28341076777 \pi^{8}}{813676722388992000}m_{6,4} + \frac{6646517 \pi^{8}}{134536495104000}m_{5,5}\\
        &+ \frac{1720681199 \pi^{10}}{2847868528361472000}m_9 + \frac{5228546699 \pi^{10}}{158214918242304000}m_{8,1} + \frac{29734883489 \pi^{10}}{65922882600960000}m_{7,2}\\
        &+ \frac{2833604759 \pi^{10}}{1255673954304000}m_{6,3} + \frac{27621933149 \pi^{10}}{5650532794368000}m_{5,4}\\
        &+ \frac{2093413439 \pi^{12}}{13184576520192000}m_8 + \frac{1028275626517 \pi^{12}}{148326485852160000}m_{7,1} + \frac{1547183158841 \pi^{12}}{21189497978880000}m_{6,2}\\
        &+ \frac{71870494697 \pi^{12}}{264868724736000}m_{5,3} + \frac{21290470169 \pi^{12}}{51368479948800}m_{4,4}\\
        &+ \frac{723245314721 \pi^{14}}{24721080975360000}m_7 + \frac{10482426936559 \pi^{14}}{10594748989440000}m_{6,1} + \frac{3415903160291 \pi^{14}}{441447874560000}m_{5,2} + \frac{278960556041 \pi^{14}}{13759414272000}m_{4,3}\\
        &+ \frac{714235157688961 \pi^{16}}{190705481809920000}m_6 + \frac{374872663008227 \pi^{16}}{3973030871040000}m_{5,1}\\
        &+ \frac{15029334826883 \pi^{16}}{28894769971200}m_{4,2} + \frac{2706543742892333 \pi^{16}}{3033950846976000}m_{3,3}\\
        &+ \frac{1273141965049441 \pi^{18}}{3910952263680000}m_5 + \frac{91605521229823 \pi^{18}}{15801827328000}m_{4,1} + \frac{424892819083031 \pi^{18}}{20316635136000}m_{3,2}\\
        &+ \frac{21003173757852809 \pi^{20}}{1137731567616000}m_4 + \frac{3833971660458281 \pi^{20}}{17777055744000}m_{3,1} + \frac{639949676749247 \pi^{20}}{1410877440000}m_{2,2}\\
        &+ \frac{1335614874712607261 \pi^{22}}{2085841207296000}m_3 + \frac{22792697163251567591 \pi^{22}}{5214603018240000}m_{2,1}\\
        &+ \frac{27265508288173067377 \pi^{24}}{2234829864960000}m_2 + \frac{103459516577811119839 \pi^{24}}{2607301509120000}m_{1,1}\\
        &+ \frac{1333969300247433001961 \pi^{26}}{12710594856960000}m_1 + \frac{158075460169843246549 \pi^{28}}{605266421760000}
    \end{aligned}
\end{equation}
}
\newpage
\printbibliography

\end{document}